# Detecting Misinformation in Multimedia Content through Cross-Modal Entity Consistency: A Dual Learning Approach

*Completed Research Paper*

**Zhe Fu**
UNC Charlotte, Software and
Information Systems
Charlotte, USA
zfu2@charlotte.edu

**Kanlun Wang**
UNC Charlotte, Business Information
Systems and Operations Management
Charlotte, USA
kwang17@charlotte.edu

**Wangjiaxuan Xin**
UNC Charlotte, Software and
Information Systems
Charlotte, USA
wxin@charlotte.edu

**Lina Zhou**
UNC Charlotte, Business Information
Systems and Operations Management
Charlotte, USA
lzhou8@charlotte.edu

**Shi Chen**
UNC Charlotte, Public Health Sciences
Charlotte, USA
schen56@charlotte.edu

**Yaorong Ge**
UNC Charlotte, Software and
Information Systems
Charlotte, USA
yge@charlotte.edu

**Daniel Janies**
UNC Charlotte, Bioinformatics and
Genomics
Charlotte, USA
djanies@charlotte.edu

**Dongsong Zhang**
UNC Charlotte, Business Information
Systems and Operations Management
Charlotte, USA
dzhang15@charlotte.edu

## Abstract

*The landscape of social media content has evolved significantly, extending from text to multimodal formats. This evolution presents a significant challenge in combating misinformation. Previous research has primarily focused on single modalities or text-image combinations, leaving a gap in detecting multimodal misinformation. While the concept of entity consistency holds promise in detecting multimodal misinformation, simplifying the representation to a scalar value overlooks the inherent complexities of high-dimensional representations across different modalities. To address these limitations, we propose a Multimedia Misinformation Detection (MultiMD) framework for detecting misinformation from video content by leveraging cross-modal entity*

*consistency. The proposed dual learning approach allows for not only enhancing misinformation detection performance but also improving representation learning of entity consistency across different modalities. Our results demonstrate that MultiMD outperforms state-of-the-art baseline models and underscore the importance of each modality in misinformation detection. Our research provides novel methodological and technical insights into multimodal misinformation detection.*

# Introduction

Social media platforms have emerged as pivotal communication channels for users to generate and share ideas, contributing to a growing volume of information dissemination within online communities (Kapoor et al., 2018). The dissemination of authentic information uplifts and nurtures a healthy online environment, fostering users' emotional well-being through the promotion of positivity, happiness, and a sense of community (Akram & Kumar, 2017; K. Wang et al., 2023). In contrast, the dissemination of falsified information can lead to a substantially adverse impact on various targets, ranging from individual users to the general public (L. Wu et al., 2019). The substantial influx of misinformation on social media platforms has sparked significant concerns across various domains, including public health (Naeem & Bhatti, 2020), business (Rangapur et al., 2023), and politics (Bovet & Makse, 2019), further deteriorating the social environment and online ecosystem. For instance, the proliferation of online misinformation on coronavirus disease in the first three months of 2020 resulted in nearly 6,000 people being hospitalized worldwide (WHO, 2021). In addition, government regulation of online video services across different continents is on the rise ("Trends and Issues in Online Video Regulation in the Americas," n.d.). For instance, the European Union has introduced new obligations on video-sharing platforms, which require platforms to safeguard minors from harmful content and protect the general public from illegal content or content promoting violence or hate speech. Therefore, the development of misinformation detection techniques plays a crucial role in curbing the dissemination of misinformation and promoting the legitimacy of online communications.

Despite the growing body of literature on online misinformation detection (Khattar et al., 2019; Y. Wang et al., 2021; Y. Wu et al., 2021), it lags behind the increasingly prevalent multimodal content on social media platforms. The latter is driven by the desire for diverse and rich expression, enhanced storytelling, increased engagement, and the accommodation of various user preferences (Shah, 2016). Specifically, Social Media Content (SMC) could incorporate text for conveying information, an image for visual effects, and audio for providing additional context or emotions, resulting in three primary modalities—text, image, and audio. The variety of SMC modalities enhances the interpretability of users' expressions, yet also poses significant challenges for misinformation detection. Recent research (Choi & Ko, 2022; Qi, Bu, et al., 2023; Shang et al., 2021) has leveraged deep learning models to extract informational cues from multimodal SMC to facilitate misinformation detection. However, the role of consistencies among multimodal SMC in misinformation detection has been severely understudied.

Inconsistency has been considered as an indicator of deception or veracity in the literature (Heinrich & Borkenau, 1998; Shan et al., 2021). An entity, typically referring to a person, location, or organization mentioned in SMC, can provide important insights for evaluating the consistency across modalities. Emerging studies have explored the role of entity consistency across different modalities in misinformation detection from multimodal SMC (e.g., Qi et al. 2023; Tan et al. 2020). However, these studies, focused on the text and image modalities, fail to address SMC with more than those two modalities. For example, as another form of medium in SMC, video is inherently more complex (e.g., a combination of text, image, and audio), as shown in Figure 1, and is widely used on video-centric social media platforms (e.g., YouTube, TikTok). As a result, it is more complex and challenging to identify entities from the video content and measure the consistency of entities between different modalities in video-centric SMC. Last but not least, existing studies simply concatenate the scalar value of cross-modal entity consistency to SMC representation, which lacks robustness and may lead to information loss.

To fill the above-mentioned research gaps, this study proposes a Multimedia Misinformation Detection (MultiMD) model enhanced by cross-modal entity consistency for SMC. It aims to answer the following research questions:

RQ1. Can the proposed model outperform the state-of-the-art multimodal models in detecting misinformation?

RQ2. Does learning a high-dimensional representation of entity consistency enhance the performance of multimodal misinformation detection? and

RQ3. How does each modality of SMC contribute to the performance of misinformation detection models?

The proposed MultiMD makes three-fold contributions to the misinformation detection literature. First, prior studies have primarily focused on either a single modality or a combined modality of text and image. In this study, we extend the scope of multimodality by incorporating text, image, and audio into the same model for misinformation detection. Second, among the limited existing studies that have explored entity consistency in misinformation detection, they represent entity consistency using scalar values, which are insufficient and lack robustness. In contrast, representing information as vectors in higher-dimensional space has shown to be more robust and informative in machine learning (Mackenzie, 2015; Mnih & Salakhutdinov, 2007). However, it requires the design of new methods to represent the cross-modal entity consistency more effectively. In this study, we introduce a dual learning approach to represent entity consistency, which encompasses two learning tasks: a primary learning task of misinformation detection and an auxiliary learning task of entity consistency measurement. In addition, we propose a hierarchical similarity-based approach to measure the cross-modal entity consistency at two different levels (i.e., modality level and SMC level) to support the auxiliary learning task.

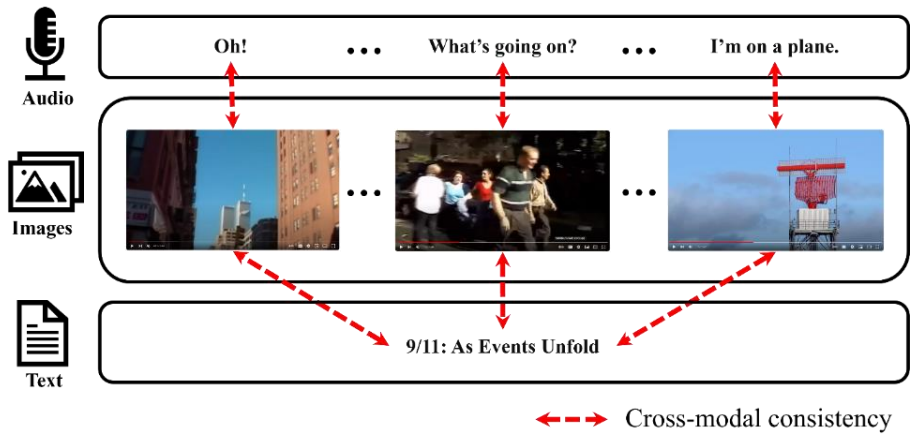

**Figure 1. Illustration of Entity Consistency in Video-centric SMC**

# Related Work

This section reviews related work on multimodal social media content and misinformation detection models, particularly deep learning-based models.

## *Multimodal Social Media Content*

Social media platforms allow users to express themselves and share information and opinions in multiple modalities, including text, image, and audio.

**Text** generally encompasses the text title, description, and other relevant textual content (e.g., captions and/or transcriptions of videos) of social media posts. Characteristics of textual content may include linguistic features (e.g., Papadopoulou et al. 2017; Volkova et al. 2017), lexicon-based psycholinguistic features (e.g., Linguistic Inquiry and Word Count (Tausczik & Pennebaker, 2010)), and vectorized features (e.g., GloVe (Pennington et al., 2014) and Word2Vec (Mikolov et al., 2013)). Moreover, there has been increasing attention to leveraging pre-trained language models (e.g., BERT (Devlin et al., 2019) and RoBERTa (Liu et al., 2019)) to obtain the latent representation of text.

**Images** play a crucial role in information sharing due to their richness and expressiveness, making them a valuable supplementary component for misinformation detection. Exemplary studies that focus on detecting visual tampering clues leverage the error level analysis algorithm (Xue et al., 2021) or image noise consistency to detect tampering traces (Mahdian & Saic, 2009). Along with others, semantic features (Cao et al., 2020), distinct fake image features (Jin et al., 2016), and emotional effects (Shu et al., 2017) have also been proven effective in capturing the nuanced context of images.

**Audio** modality involves acoustic signals or voice waves, which are conveyed from speech, environment sounds, and background music (Bu et al., 2023). While audio is always integrated into videos, it also exists independently in various forms, such as podcasts and audio files (Comito et al., 2023). To leverage the audio modality, Hou et al. (2020) utilized raw low-level acoustic features with the openEAR (Eyben et al., 2009) toolkit, and Shang et al. (2021) transformed audio segments into fix-length embedding vectors by extracting the Mel-frequency Cepstral Coefficients features (Palo et al., 2018). The emergence of pre-trained models like VGGish (Hershey et al., 2017) has sparked widespread use for extracting audio content features in recent studies (Lin et al., 2023; Mao et al., 2023; Qi, Bu, et al., 2023).

### Deep Learning-Based Misinformation Detection Models

In recent years, there has been a surge of research on social media analytics and online misinformation detection, driven by the remarkable effectiveness and efficiency of deep learning (e.g., Choi and Ko 2022; Khattar et al. 2019; Qi et al. 2023; L. Wang et al. 2023).

Previous studies have primarily concentrated on the text modality for misinformation detection (Guacho et al., 2018; Sarin & Kumar, 2020; L. Wu et al., 2019). The primary focus of those studies was on leveraging plain text to detect misinformation, often entailing the training of a classifier using textual posts sourced from public SMC. Due to the diverse nature of SMC disseminated on social platforms, there has been increasing attention to detecting misinformation from SMC in text and/or image modalities. Previous research (Khattar et al., 2019; Y. Wang et al., 2021; Y. Wu et al., 2021) focused on the use of co-attention mechanisms to fuse the features of text and image content. For instance, Z. Wang et al. (2021) incorporated semantic- and task-level attention mechanisms to enhance deep learning model performance to detect anti-vaccine-related misinformation from medical posts on Instagram. Y. Wu et al. (2021) employed multimodal co-attention mechanisms to facilitate misinformation detection by stacking co-attention layers after extracting features from images and textual content. Khattar et al. (2019) explored the use of autoencoders to compress pertinent cues in text-image modalities. In addition to studies solely dependent on extracted features of multimodal content to improve model performance, some endeavors have been undertaken to incorporate external knowledge into detection models. For example, Qian et al. (2021) proposed a knowledge-aware multimodal adaptive graph convolutional network framework, which represented social media posts as graphs and integrated external knowledge (i.e., conceptualized entity relations) from a knowledge graph constructed through a set of text entities. However, their approach of fusing features through a basic concatenation operation, incorporating image representations from VGG and text presentations from a graph convolutional network framework, proved inadequate in capturing intricate inter-modality relationships.

Y. Wang et al. (2021) detected fake news related to emergent events using an end-to-end meta-neural process framework with a limited set of verified posts to explore the modality correlation between textual and visual news content. Similarly, Qi et al. (2021) proposed an entity-enhanced framework that leveraged three text-image correlations, including entity consistency, mutual enhancement (i.e., the text-image semantic relation), and text complementation. Chen et al. (2022) proposed and investigated the concept of cross-modal ambiguity, which was measured by Kullback Leibler (KL) divergence between text and image feature distributions. Nevertheless, this study lacks comprehensiveness as the ambiguity measurements concentrate on the high-level representations (e.g., embedding vector-level) while ignoring entity-level representations. In contrast, Zhou et al. (2020) introduced a Similarity-Aware FakE (SAFE) news detection model, learning cosine similarity-based relationships between images and text. Notably, in their research, features of images were generated from their textual captions, which were converted from raw images by using an image-to-sentence tool (Vinyals et al., 2017). This incurs potential information loss due to the limitations of the text corpus used by the tool.

Despite the significant research in text- and/or image-based misinformation detection, video-centric SMC remains underexplored for misinformation detection. Among the few limited studies, Choi and Ko (2022) leveraged domain knowledge (i.e., domain-specific features related to fake news) to validate the legitimacy of video-centric SMC. Shang et al. (2021) investigated COVID-19-related short videos on TikTok by extracting features across different modalities and aggregating heterogeneous clues. Qi, Bu, et al. (2023) introduced a Chinese short video dataset and a video misinformation detector. In addition, some others have applied both supervised- and self-supervised learning methods (K. Wang, Chan, et al., 2022) and contrastive learning, masked language modeling, and/or cross-sampling (Qi, Zhao, et al., 2023; X. Wu, 2023) to detect misinformation.

In summary, previous research in misinformation detection has primarily focused on either a single modality or the fusion of text and image modalities. Studies addressing text, images, and audio remains scarce. In addition, while entity consistency has demonstrated potential as an indicator for misinformation detection (e.g., Qi et al. 2023; Tan et al. 2020), current measures of entity consistency rely on a scalar value derived from various modalities during model development without actually learning the consistency among multimodal content or entities.

## Method

We propose MultiMD, a cross-modal entity consistency-based approach via dual learning to determine whether video-centric SMC is fake (i.e., misinformation) or not. The approach utilizes entity consistency across three different modalities to enhance model performance. MultiMD consists of two novel artifacts for multimodal fusion (see Figure 2): a dual learning mechanism to learn a high-dimension representation of entity consistency from the multimodal representations of SMC and a hierarchical computational approach to measuring entity consistency in video-centric SMC.

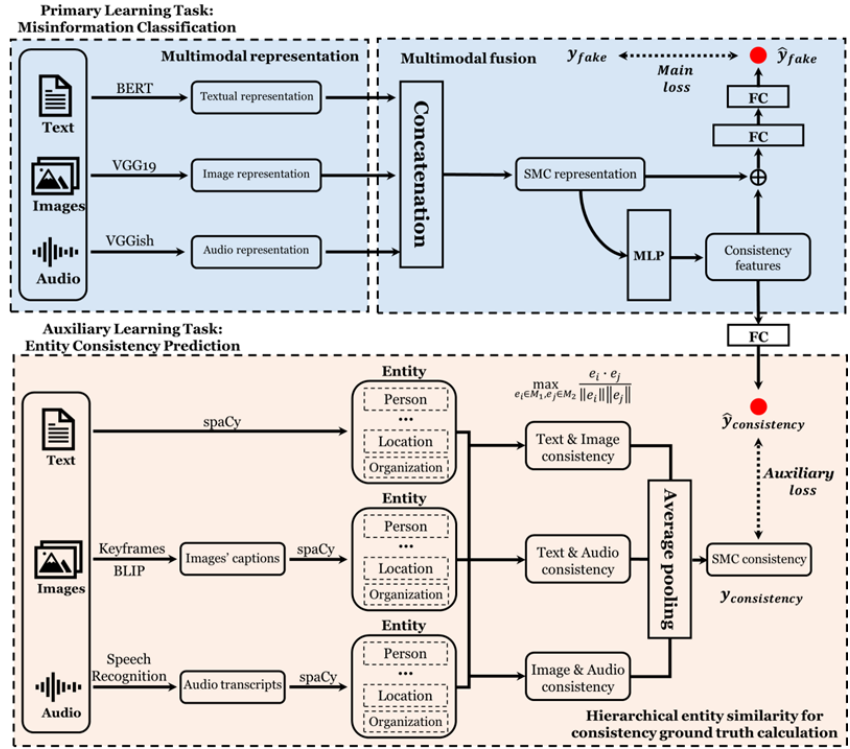

**Figure 2. The Framework of MultiMD**

### Problem Formulation

We frame misinformation detection as a binary classification problem, wherein SMC in the form of a video is classified as either real or fake. We represent SMC as $S = \{T, I, A\}$, where $T$, $I$, and $A$ denote textual, image, and audio content, respectively. To effectively learn an SMC representation, we aggregate the video content in three modalities (i.e., text, image, and audio) in building a classifier, as shown in equation (1).

$$y = f(\theta, S) \tag{1}$$

Where $f(\cdot)$ is the classification function; $y$ denotes the classification result of $S$; and $\theta$ denotes the set of parameters of $f(\cdot)$.

### Multimodal Content Representation

#### Text

To represent the textual content, including the titles and descriptions of SMC, we use a pre-trained BERT embedding model (Devlin et al., 2019) to encode the text into vector representations, given that the effectiveness of BERT encoder has been well demonstrated by previous NLP studies (Kaliyar et al., 2021; Minaee et al., 2021; K. Wang, Fu, et al., 2022). We deploy a special classification token (i.e., $[CLS]$) at the beginning of a word sequence and the ultimate hidden state associated with this token serves as the consolidated sequence representation. We obtain the representation of the textual content $h_T$ with a dimension size of 768.

#### Images

For image representation, we first conceptualize the image content $I = \{F_1, F_2, \dots, F_k\}$ as a sequence of video frames, where $F_k$ is an image frame in a video file and $k$ represents the total number of frames sampled at 1-second intervals from the video. Drawing on prior studies (e.g., Khattar et al. 2019; Qi et al. 2021), we adopt VGG19 (Simonyan & Zisserman, 2014), a CNN-based model pre-trained on ImageNet, to generate the representation of each frame in the image content. To generate a unified representation of image content $h_I$, we aggregate the 1,024-dimension representation vectors of all video frames with an average pooling mechanism (see Equation 2).

$$h_I = ave(\,[\,F_1, F_2, \dots, F_k\,]) \tag{2}$$

#### Audio

To represent audio content (e.g., the soundtrack associated with the video), we segment the audio content of video file $A$ into a sequence of consecutive chunks $A = \{C_1, C_2, \dots, C_k\}$, where $C_k$ is an audio chunk in $A$ and $k$ is the total number of chunks sampled at 1-second intervals from $A$. It is worth noting that the sampling frequencies of image and audio content are synchronized. Following previous studies (e.g., Chen and Zhang 2023; Qi, Bu, et al. 2023), we deploy the VGGish model (Hershey et al., 2017) to extract wave signals of audio content effectively. Finally, we generate a unified representation of the audio, $h_A$, through average pooling of the 128-dimension vectors of all audio chunks.

### Multimodal Fusion

We concatenate the representation vectors of all three modalities, as shown in Equation 3.

$$h_{SMC} = [\,h_T;\ h_I;\ h_A\,] \tag{3}$$

Where $h_T \in R^{1 \times 768}$, $h_I \in R^{1 \times 1024}$, and $h_A \in R^{1 \times 128}$ are the representation vectors of the three modalities.

### Dual Learning

Despite the benefit of concatenating multimodal representations in the SMC representation for information retention without incurring substantial information loss, it might introduce noise that can distract the model's attention and compromise the reliability of prediction results (H. Li et al., 2018). To address this issue, a few studies have explored entity consistency across different modalities as crucial clues to enhance the effectiveness of misinformation detection (e.g., Qi et al. 2023; Tan et al. 2020). However, these studies fail to address SMC with more than two modalities. More importantly, existing studies simply leverage the scalar value of cross-modal entity consistency, which lacks sufficient information and robustness.

To address the above limitations, we introduce a dual learning approach. The approach consists of two learning tasks, with the primary task focusing on misinformation detection and the auxiliary learning focusing on measuring cross-modal entity consistency. The primary learning task aims to obtain an effective SMC representation for misinformation detection, while the auxiliary task aims to learn a high-dimensional representation for entity consistency. By performing the two learning tasks simultaneously, the proposed dual learning approach can mutually optimize the representations of both SMC and entity consistency and consequently enhance the performance of MultiMD.

**Primary Learning Task: Misinformation Classification**

With the generated representation of SMC $h_{SMC}$, we build a classification model by designing two fully connected layers with a SoftMax function to generate the probability distributions for the two target classes — fake and real (see Equation 4).

$$\hat{y} = softmax(W_2\,\phi(W_1 h_{SMC} + b_1) + b_2) \tag{4}$$

where $\hat{y}$ is the classification result of SMC. $W_1 \in R^{d_1 \times d}$ and $W_2 \in R^{2 \times d_1}$ denote the learnable weights; $d$ is the dimension of SMC representation $h_{SMC}$; $d_1$ is the dimension of the hidden output of the first fully connected layer; $b_1$ and $b_2$ are the biases for the respective fully connected layers; and $\phi$ is the activation function (e.g., sigmoid, Tanh, and ReLU). We choose the binary cross-entropy as the loss function (see Equation 5), where $N$ denotes the training data size. The learning goal is to minimize the loss value.

$$L(\theta) = -\frac{1}{N}\sum_{i=1}^{N}[y_i log(\hat{y}_i) + (1 - y_i)log(1 - \hat{y}_i)] \tag{5}$$

**Auxiliary Learning Task: Entity Consistency Estimation**

Previous studies suggest that cross-modal consistency can serve as a potential indicator for misinformation detection (e.g., Müller-Budack et al. 2020; Qi et al. 2021; Sun et al. 2023). Therefore, we anticipate that incorporating entity consistency into the SMC representation would further enhance the performance of our primary learning task. To this end, we propose a Multi-Layer Perceptron (MLP) as a feature extractor to extract entity consistency features $h_{consistency}$ from the SMC representation.

$$h_{consistency} = MLP(h_{SMC}) \tag{6}$$

To empower the extractor with the ability to learn useful features for inferring cross-modal consistency, we design another auxiliary learning loss function for the cross-modal consistency prediction task (see Equation 7).

$$L_{aux} = [\,(W_c h_{consistency} + b_c) - y_{consistency}\,]^2 \tag{7}$$

where $W_c \in R^{1 \times d_c}$ denotes learnable weights; $d_c$ is the dimension of $h_{consistency}$; $b_c$ is the bias for the fully connected layer, and $y_{consistency}$ is the ground truth of cross-modal consistency.

Given the absence of ground truth for entity consistency in our dataset, we propose a hierarchical entity similarity-based approach to generate pseudo ground truth for the auxiliary learning task automatically. Specifically, we extract entities from the textual, image, and audio content of SMC separately.

- *Textual content*. We leverage spaCy[1], a software library for advanced natural language processing, to extract named entities (e.g. persons, locations, organizations) from the textual content of SMC. Then, we use the pre-trained word2vec[2] to encode the extracted entities as embedding vectors.
- *Image content*. We first extract the keyframes of each video, considering that a significant portion of frames in the same video are likely to be identical or similar to one another with minimal changes between consecutive scenes. We then leverage the BLIP model (J. Li et al., 2022), a vision-language pre-trained model, to caption each keyframe of a video with a short phrase. To obtain entities from the image content, we aggregate the entities extracted from the caption of each keyframe and employ the method for textual content to encode the identified entities.

---

[1] https://spacy.io/usage/linguistic-features/
[2] https://radimrehurek.com/gensim/models/word2vec.html

- *Audio content.* We first transcribe it into text using the SpeechRecognition tool[3], a library supporting speech recognition. Subsequently, we extract entities from text to the audio transcription.

To account for the contradictory content across different modalities that misinformation might incorporate, we develop measures for cross-modal consistency based on entity similarity. Moreover, we measure the hierarchical consistency at two different levels: modality and SMC. At the modality level, we measure the entity consistency between any pair of modalities based on the maximum consistency value, as shown in Equation (8).

$$consistency_{M_1, M_2} = \max_{e_i \in M_1, e_j \in M_2} \frac{e_i \cdot e_j}{\|e_i\| \|e_j\|} \qquad (8)$$

where $e_i$ and $e_j$ are two entity embedding vectors extracted from modalities $M_1$ and $M_2$, respectively.

At the SMC level, we aggregate the three modality-level consistency values (i.e., text-image, text-audio, and image-audio consistency) in video-centric SMC by averaging them. We use the SMC-level entity consistency as the pseudo ground truth for consistency in the auxiliary learning task. Finally, we enhance the representation of $h_{SMC}$ by concatenating it with the extracted feature vector $h_{consistency}$, as shown in Equation (9).

$$h'_{SMC} = [\, h_{SMC}; h_{consistency} \,] \qquad (9)$$

The enhanced SMC representation $h'_{SMC}$ learned by the auxiliary learning task is expected to improve the performance of the primary learning task (i.e., the misinformation classification task formulated in Equation 4). Meanwhile, the auxiliary learning task can learn a more informative and robust representation of entity consistency with a more effective SMC representation learned through the primary task.

# Evaluation

## Dataset and Preparation

We collected four benchmark misinformation datasets, which contain multimodal SMC from YouTube and are in the form of text and video, including YouTubeAudit (Hussein et al., 2020), FVC (Papadopoulou et al., 2018), VAVD (Palod et al., 2019), and NLP-COVID (Serrano et al., 2020). Given that these datasets only contain the URLs but not actual video content, we first extracted the URLs of video files from the SMC and then utilized YT-DLP[4] to scrape the corresponding video content from the YouTube platform. Data collection was confined to publicly available posts to adhere to YouTube's privacy policy. Consequently, the data collection procedure did not require a pre-approval from the Institutional Review Board.

To enhance the ecological validity of the study findings, we pre-processed the benchmark datasets by removing redundant URLs and irrelevant SMC (e.g., without a video title or in a non-English language). Given that the data with fake and real labels were imbalanced in the collected datasets, a common problem in classification tasks that may result in poor model performance, we performed under-sampling on the real data, resulting in 1,288 SMC instances equally divided between the fake and real categories.

## Baseline Models

We selected the following four state-of-the-art models for multimodal misinformation detection as baselines in this study.

---

[3] https://github.com/Uberi/speech_recognition
[4] https://github.com/yt-dlp/yt-dlp

- EM-FND (Qi et al., 2021) uses textual and visual features extracted by BERT and VGG19 as base features and the value of entity consistency as an additional feature. It calculates consistency by averaging the three cross-modal consistency values.
- MVAE (Khattar et al., 2019) uses a variational autoencoder to fuse the representations of different modalities.
- MCAN (Y. Wu et al., 2021) uses stacked co-attention networks to fuse the representations of different modalities.
- FVDM (Choi & Ko, 2022) is a video-based misinformation detection model that encodes video by the attention output between video frames and corresponding thumbnails. Given that there is no user comment in our dataset, we removed the comment encoding component from the model.

It is worth noting that very few existing models for multimodal misinformation detection from SMC include text, image, and audio content. For instance, EM-FND, MVAE, and MCAN only employed text and image modalities. In this study, we extended those models to address three modality scenarios.

### *Evaluation Metrics and Settings*

We employ a set of widely used evaluation metrics to assess the model performance, including accuracy, precision, recall, and F1-measure. Precision measures the proportion of detected fake SMC actually being fake; recall measures the proportion of actual fake SMC that is detected correctly; F1 is a harmonic mean of precision and recall (i.e., $\frac{2*Precision*Recall}{Precision+Recall}$); and accuracy is measured as the ratio of the sum of true positives and true negatives to all the predictions.

We deploy 10-fold cross-validations by performing a 90/10 data split for training and testing. We report the results averaged across the 10 folds. In addition, we use the learning rate of 0.01; the dimension of each fully connected layer in the classifier and that of the consistency feature extractor is 1,024; and the dropout rate is 0.2 for all the models. All of the models are optimized using the Adam optimizer.

## Results

### *Misinformation Detection Performance*

We evaluate the effectiveness of our proposed model for misinformation detection by comparing its performance with those of the baseline models. The model performances are reported in Table 1 and the results of paired sample t-tests are presented in Table 2. Notably, MultiMD demonstrates significantly higher accuracy ($p<.001$ or $p<.01$) and F1 scores ($p<.001$) than all baseline models. In addition, MultiMD demonstrates significantly higher recall than EM_FND ($p<.01$), MVAE ($p<.001$), MCAN ($p<.001$), and FVDM ($p<.05$) models. Furthermore, MultiMD achieves significantly higher precision than MVAE ($p<.001$), and marginally higher precision than the EM_FND, MCAN, and FVDM models ($p<.1$).

| Table 1. Model Performances of the MultiMD and the Baseline Models | | | | |
|---|---|---|---|---|
| **Models** | **Accuracy** | **Precision** | **Recall** | **F1** |
| EM_FND | 65.38% | 66.11% | 67.08% | 64.66% |
| MVAE | 62.99% | 63.23% | 64.47% | 62.88% |
| MCAN | 64.07% | 66.00% | 58.21% | 60.10% |
| FVDM | 64.90% | 64.43% | 72.10% | 65.97% |
| *MultiMD* | **74.07%** | **71.35%** | **83.75%** | **76.08%** |
| *Note: The best results are highlighted in boldface.* | | | | |

| Table 2. Performance Comparison between the MultiMD and the Baseline Models | | | | |
|---|---|---|---|---|
| **Models** | **Accuracy** | **Precision** | **Recall** | **F1** |
| EM_FND | -5.667*** | -2.169[†] | -4.084** | -7.044*** |
| MVAE | -6.227*** | -5.861*** | -6.141*** | -7.322*** |
| MCAN | -5.156*** | --1.980[†] | -7.633*** | -7.276*** |
| FVDM | -3.979** | -2.217[†] | -2.710* | -4.787*** |
| *t-statistic: ***: p<.001; **: p<.01; *: p<.05; [†]: <.1.* | | | | |

### *Ablation Experiments*

To examine the impacts of different modalities and cross-modal entity consistency on model performance, we further conducted an ablation experiment by removing one modality or the consistency measure (i.e., auxiliary learning task) from MultiMD and performed multiple paired samples t-tests.

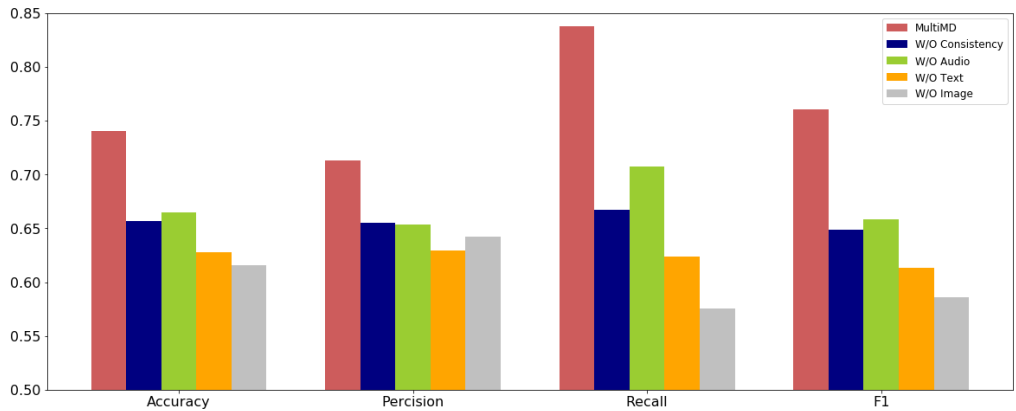

**Figure 3. Ablation Experiment Results**

| Table 3. Performances Comparisons of the Ablated Models vs. the Full MultiMD Model | | | | |
|---|---|---|---|---|
| **Ablated Modality** | **Accuracy** | **Precision** | **Recall** | **F1** |
| Image | .125** | .071** | .262*** | .175*** |
| Text | .113*** | .084** | .214*** | .148*** |
| Audio | .075** | .060[†] | .130** | .103*** |
| Consistency | .084*** | .059* | .170** | .112*** |
| *Note: When Image is the ablated modality, it refers to the MultiMD model with the image modality removed. It is the same for the other three ablated models.*<br>*** : p<.001; **: p<.01; *: p<.05; [†]: <.1.* | | | | |

The results of the ablation experiment are presented in Figure 3 and the paired t-test comparison results along with their t-statistics are reported in Table 3. Each numeric value in Table 3 represents the mean difference in the corresponding performance measure between an ablated model and the full MultiMD model. The results show that MultiMD achieves the optimal performance when combining all three modalities along with the entity consistency measure. In addition, each of the three modalities contributes to detection performance, which is manifested by the decrease in model performance across all four performance metrics after removing individual modalities from MultiMD. Among the different modalities, ablating the image modality leads to the most significant performance drop across all the performance measures, followed by the text modality, indicating that image content plays the most important role in multimodal misinformation detection, whereas the audio content contributes to the least.

## Discussion and Conclusion

As the expression of misinformation becomes richer, the method for its detection should leverage multimodal information to capture a broader range of deceptive tactics. This study investigates an understudied area of misinformation detection by developing a novel multimodal framework for detecting misinformation from video-centric SMC. The framework not only expands the scope of literature by integrating text, image, and audio content for misinformation detection, but also introduces a dual learning-based method for enhancing video representations. In particular, the dual learning method encompasses a primary task centered on misinformation detection alongside an auxiliary task focused on measuring entity consistency.

In response to RQ1, our empirical evaluation results demonstrate that the model based on our proposed framework (i.e., MultiMD) outperforms the state-of-the-art multimodal misinformation detection models across all evaluation metrics. Notably, our proposed entity consistency mechanism significantly augments the model's overall performance. The results provide a strongly positive response to RQ2. Furthermore, in response to RQ3, our results show that the model incorporating all three modalities leads to the best performance. In other words, each modality of the video-centric SMC contributes to the model performance.

This research offers opportunities for future extensions. First, this study mainly focuses on learning useful information from SMC, such as representations of individual modalities and entity consistency. Our proposed framework can be further enhanced by incorporating the importance of individual modalities and/or leveraging external knowledge sources to facilitate misinformation detection. Second, our proposed framework leverages various tools for preprocessing multimodal data, such as entity extraction and image captioning tools. These tools may introduce noise to the dual learning process. Similarly, we propose a novel method for generating the studio ground truth for entity consistency, which however may not perfectly mirror cross-model consistency. Thus, the performance of our framework can benefit from future advances in these related techniques. Third, while we derive consistency measures based on the entities in SMC, other forms of information such as sentiment can serve as alternative bases for measuring consistency, warranting further investigation in future research. Moreover, the finding about the positive impact of cross-model consistency on the performance of misinformation detection models opens a new avenue for future research. For instance, future research can delve into the role of consistency in misinformation detection, offering deeper insights into how the consistency measure interacts with individual modalities to bolster the model performance. Last but not least, an examination of our proposed framework for classification tasks across different problem domains employing video-centric SMC would contribute to a broader understanding of its generalizability. One interesting extension of this research is to evaluate the efficacy of the framework in identifying misinformation from AI-generated video content.

## Acknowledgments

This work is partially supported by a Truist Research Grant. Any opinions, findings, or recommendations expressed here are those of the authors and are not necessarily those of the sponsors of this research.